\documentstyle[psfig]{kluwer}

\runningtitle{The Dark Matter problem}
\runningauthor{A. Bosma}

\begin{opening}

\title{The DARK MATTER problem}

\bigskip
\author{Albert \surname{Bosma}}

\institute{Observatoire de Marseille\\
 2 Place Le Verrier, 13248 Marseille C\'edex 4, FRANCE}

\end{opening}

\begin{document}

\font \usualfont=cmr10
\font \titlefont=cmbx10
\font \smallfont=cmr9
\font \smallbold=cmbx9
\font \subsecfont=cmr8
\font \smallitalic=cmti9
\font \runningfont=cmr6

\def\tab{\hspace*{0.8truecm}}
\def\ref{\par\noindent\hangindent 20pt} 
\def \ms{\medskip \noindent}
\def\@cite#1#2{{#1\if@tempswa , #2\fi}}

\begin{abstract}
In these notes I will briefly summarize our knowledge about the dark
matter problem, and emphasize the corresponding dynamical aspects.
This covers a wide area of research, so I have been selective, and
have concentrated on the subject of dark matter in nearby galaxies, 
in particular spirals.
\end{abstract}


\bigskip
\usualfont

\section{Scope of the dark matter problem}

\subsection{Several dark matter problems ?}

There are several dark matter problems, each on a different scale. 
The standard way to prove the existence of dark matter is to compare 
dynamical estimates of the mass of an object (galaxy, cluster) with an 
evaluation of its population contents. Historically, we can
distinguish four distinct problems : \\
\indent 1. the galactic force law in the solar neighbourhood \\
\indent 2. galactic halos \\
\indent 3. groups and clusters of galaxies \\
\indent 4. the Universe as a whole.

\noindent
In each case there is more matter inferred dynamically than can
be accounted for by known matter components. This mass discrepancy 
is usually attributed to additional (dark) matter, assuming that
Newton's laws are
valid. Only in Modified Newtonian Dynamics (MOND, cf. Milgrom,
1983) is the discrepancy attributed to a modification of 
the force law at low densities. This theory has been worked out
in detail only for the explanation of spiral galaxy rotation curves.
The evidence for dark matter from extended rotation curves of
spiral galaxies is considered the strongest, and this topic has thus
spawned the most adhoc alternatives. Not only MOND, but also 
explanations based on magnetic forces have been considered. (cf. 
Battaner 1992, but see Cuddeford \& Binney 1993). For a review of 
alternative theories of gravity, see Sanders (1990).

In these lecture notes, I will only discuss certain aspects of the
whole dark matter problem, with a distinct emphasis on dark matter 
in spirals, including the Milky Way. The lectures were given to an
audience of scientists working mainly in celestial mechanics, hence
the inclusion of a lot of introductory material.

\subsection{Dark matter in the solar neighbourhood}

Oort (1932, 1960) was the first to perform an 
analysis of the vertical equilibrium of the stellar distribution in 
the solar neighbourhood. He found that there is more mass in the
galactic disk than can be accounted for by star counts. A reanalysis
of this problem by Bahcall (1984abc) argued for the presence of a dark
``disk'' of a scaleheight of 700 pc. This was called into question
by Bienaym\'e, Robin \& Cr\'ez\'e (1987), and by Kuijken \& Gilmore 
(1989ab, 1991). 
The newest result is based on a sample of stars with HIPPARCOS
distances and Coravel radial velocities, within 125 pc of the Sun.
Cr\'ez\'e et al. (1998) find that there is no evidence for dark matter
in the disk : all the matter is accounted for by adding up the 
contributions of gas, young stars and old stars.

\subsection{Dark matter in groups and clusters}

An analysis of 21 radial velocities of galaxies in the Coma
cluster, assuming virial equilibrium, led Zwicky (1933) to 
conclude that the integrated mass-to-light ratio ${\Upsilon}$ of 
those galaxies must be of order 100 - 500. This result has
been confirmed in more recent studies of clusters of 
galaxies (cf. sec. 4.2).

A classical paper on dark matter in the Local Group is Kahn \& Woltjer
(1959). Several notions still of interest are introduced in this paper.
One is on the persistence of warps. A warp in the outer HI layer of the
Milky Way had been discovered by the work of Burke (1957) and Kerr 
(1957). Kahn \& Woltjer note that for any non-spherical distribution of
mass a warp will suffer from differential precession. Thus a
tidally pulled warp, with an integral sign warp shape, will wrap 
up and corrugate the disk in at most a few Gyr. It is clear that warps
suffer from a similar persistence problem as do spiral arms.

The most interesting argument in Kahn \& Woltjer (1959) is 
called the ``timing
argument'', and is based on the notion that M31 approaches our Galaxy
with a relative speed of 119 km s$^{-1}$. Such a local deviation from
the Hubble flow can only be due to the presence of mass in the Local
Group. If M31 approaches us for the first time, a total mass for the
Local Group of $\sim$ 3 10$^{12}$ M$_{\odot}$ follows. In principle
the mass could reside in each of the two principal galaxies, or else
it could be much more spread out in the entire volume of the Local 
Group. In fact Kahn \& Woltjer (1959) 
propose a much more spread out distribution in gaseous
form, and study the potential effect of magnetic forces due to an
intergalactic wind blowing on the outer gas layer of our Galaxy
thereby causing the warped shape. (The notion that gravity is the 
dominant force in the dynamics of galaxies, including interactions
between them, was firmly established 
only with the article of Toomre \& Toomre (1972) on galactic
bridges and tails).

\subsection{Dark matter in spiral galaxies}

\subsubsection{Expectations}

The notion of dark matter in and around galaxies was only gradually
developed, and became accepted during the 1970's. This slow
development is due to the complexity of the problem : our
knowledge of stellar populations in galaxies drastically
improved due to new observational techniques and more 
systematic studies. Furthermore, the development of
numerical simulations provided two very important concepts :
the idea that dynamically hot matter is needed to stabilize
disks (Ostriker \& Peebles 1973), and the notion of merging
of galaxies due to mutual interaction (Toomre \& Toomre 1972,
Toomre 1977).

If
all the mass of a galaxy is in the center, and the ionized gas can
be considered as orbiting test particles, one expects a Keplerian 
rotation curve (rotation velocity V$_{\rm rot}$ proportional to radius
R$^{-{1 \over 2}}$). If the mass distribution follows the light
distribution, i.e. the mass-to-light ratio ${\Upsilon}$ is constant
with radius, then, for an exponential light distribution (cf.
Freeman 1970) one has a rotation curve which peaks at about 2.2
times the scalelength of the disk, and which thereafter slowly
falls off towards the Keplerian behaviour.
Note that for a rotation curve that just peaks at the edge of the
optical disk, there is only 64\% of the mass enclosed inside the
radius of the disk. Hence a substantial fraction of mass is already
outside the optical radius in that case. If no turnover is reached,
more mass must be outside. This is the justification for the statement
in Freeman (1970) that for M33 and NGC 300 a lot of mass must be
outside the last measured point on the rotation curve.

\subsubsection{Early debates}

Measurements with single dish telescopes with (for that time) very
sensitive receivers provided the first hints that rotation curves
stay flat at large radii. In particular, there was a debate about the
rotation curve of M31 at large radii (cf. Roberts 1975, Emerson \& Baldwin
1973, and Baldwin 1975), but also about possible sidelobe effects
of Arecibo data (see discussion after Salpeter, 1978). Interferometer
data free from this effect for 5 Scd galaxies (Rogstad \& Shostak
1972) showed that rotation curves did not decline. Newer data for
a number of galaxies observed with the Westerbork telescope settled
these issues : a compilation of 25 rotation curves of spiral galaxies of
various morphological types showed that all of them are roughly flat,
or rising (cf. Bosma 1978, 1981a,b).

Numerical simulations of spiral galaxies also started in earnest in
the early 1970s. Hohl (1971) found consistently that flat disks
are very prone to the bar instability. A cure was devised by Ostriker
\& Peebles (1973), i.e. to embed a disk in a dynamically hot dark
halo, which was presumed to be roughly spherical.
The required halo masses interior to the disks are rather
large, so that the total mass of a large galaxy at large radii can
be easily 10$^{12}$ M$_{\odot}$. The absence of a decline in a 
rotation curve implies that mass increases linearly with radius.
In this way a mass radius relationship can be established, e.g. for
our Galaxy, using various tracers of the mass like outlying globular
clusters, satellites, Local Group timing, etc. (Ostriker, Peebles
\& Yahil 1974).

The data in the late seventies from HI observations (Bosma 1978),
show that extended flat rotation curves are ubiquitous for large spirals,
and that only for small Sc galaxies the rotation curves are still
rising. In a series of papers by Rubin et al. (1978, 1980, 1982, 1985)
on H$\alpha$ rotation curves a nice systematic behaviour as function
of type and luminosity of the rotation curves for Sa, Sb and Sc
spirals was established as well.

\subsubsection{Mass modelling methods}

Early ways of dealing with rotation curves were to take the observed
form, and to ``invert'' it to derive a density law for the disk. Such
a method is only valid if the mass resides indeed in the disk. 
If that is the case, for extended HI rotation curves, two major 
conclusions can be drawn : the mass-to-light ratio increases to very
high numbers in the outer parts, and the ratio of HI gas mass to total
mass stays roughly constant. (cf. Bosma 1978, Bosma \& Van der Kruit
1979). In the early 70s it was not yet accepted that low mass stars 
(red dwarfs) could not account for the bulk of the mass in the disk;
in the 90s the idea that dark matter is baryonic and residing in the 
disk resurfaced in the form of very cold gas, most of it undetectable 
(Pfenniger \& Combes 1994). This idea was directly based on the constancy
of the ratio of HI gas mass to total mass in the outer parts.

The inversion method leads to models which cannot be further analyzed.
A more fruitful alternative is thus to consider a ``realistic'' mass
distribution of disk and bulge, and to attribute the rotation curve
discrepancy in the outer parts to an extended dark halo. Apart from
specifying the bulge/disk decomposition, such a method requires a 
postulate concerning the mass-to-light ratio of the disk (and the 
bulge). An early 
application of this way of modelling was done by Kalnajs (1983),
who demonstrated that no dark halo is necessary when the rotation
curve does not extend far enough. As a rule, HI data extend at least
twice as far as the ``easy visible disk'', and far enough to establish
the discrepancy between the expected and the observed rotation curve,
but optical (H$\alpha$) data usually do not extend far enough to reach this
conclusion. This was further established by Kent (1986, 1987, 1988).

The question of disk stability in the presence of dark halos has
been pursued vigourously. In particular, the competing influence
of velocity dispersions in the disk on suppressing the bar instability
has been adressed by Athanassoula \& Sellwood (1986), who find that
both a massive dark halo and high disk velocity dispersions slow
down the developement of a bar. Nevertheless,
it is the initial velocity dispersion in the disk which determines
the axial ratio of the bar (Athanassoula 1983). Toomre (1981) examined the
question of spiral structure, and identified a mechanism called
swing amplification, which could lead to disks with strong
spiral structure. The presence of a dark matter halo is to lessen
the dynamical influence of the disk, and for small disk/halo ratios
the amplification may be suppressed altogether. Athanassoula, Bosma
\& Papaioannou (1987) have used this theory to try to get limits on 
the possible mass-to-light ratios for the disk.
 
\subsection{Importance of the dark matter problem today}

The dark matter problem is nowadays a complex of problems, which is 
pervasive in every aspect of extragalactic astronomy and cosmology. 
This can be gleaned from previous reviews and symposia, e.g. Trimble 
1987, Kormendy \& Knapp 1987, Holt \& Bennett 1995, Zaritsky 1998). 
Another way of seeing this is to look at the way the apparently simple
problem of the origin of the Hubble sequence has evolved. The classic 
view : ellipticals and bulges form quickly, and spirals disks are 
built up gradually, still has its supporters (e.g. Sandage 1986).
However, new ideas
about the importance of secular evolution modifying the Hubble type
of spirals, and producing ellipticals from merging disk galaxies 
(Toomre 1977), are now used together with numerical simulations of dark 
matter mixed in with gas and star formation recipes destined
to model the formation of galaxies in a cosmological framework
(e.g. Navarro, Frenk \& White 1996, 1997).
Although there are still a plethora of assumptions necessary to 
proceed from numerical simulations to observations which can test 
them, such an integrated approach seems to hold promising keys for 
future developments.

\section{Dark matter in the Galaxy}

New data at virtually all wavelength bands are now available for 
our Galaxy. In terms of the mass distribution, the most valuable
contributions come from a near IR map obtained with the COBE
satellite (Dwek et al. 1995), and the ongoing microlensing
experiments (MACHO, EROS, OGLE, etc.). For the solar neighbourhood
dynamics, the new data from the HIPPARCOS satellite will provide
fresh insights into old problems. For the determination of
the Galactic rotation curve beyond the solar radius, see the
reviews by Fich \& Tremaine 1991, Merrifield 1993, and Olling \& 
Merrifield 1998. In general, it is assumed that the Galactic rotation
curve remains more or less flat at large radii, consistent with
data on satellites such as outlying globular clusters and dwarf
spheroidals. Thus the rotation curve of our Galaxy is similar to
that of any other larger spiral.

\subsection{Escape speed for a simple spherical model}

We can tie in local observations of high velocity stars with
the notion of the escape speed (cf. Binney \& Tremaine 1987).
For a simple spherical model we have :

\begin{equation}
\left. \begin{array}{l}
M(r) = \frac{V_c^2r}{G} \hspace{0.63truecm} \mbox{at} \quad  r < r_* \\
\\
M(r) = \frac{V_c^2r_*}{G} \quad \mbox {~at} \quad r \ge r_*
\end{array}\right.
\end{equation}

\noindent
The corresponding escape speed is

\begin{equation}
\left. \begin{array}{l}
V_e^2 = 2V_c^2 [1 + \ln{r*\over r}] \quad \mbox{at} \quad r < r_* \\
\\
V_e^2 = 2V_c^2 { r_*\over r}  \hspace{1.62truecm} \mbox{at} 
\quad r \ge r_* \\
\end{array}\right.
\end{equation}

\noindent
Substituting V$_c$ = 220 km s$^{-1}$, r = R$_{\odot}$ = 8.5 kpc
and an estimate for the escape speed of 500 km s$^{-1}$, we find
r$_*$ = 4.9 R$_{\odot}$ $\simeq$ 41 kpc, and M(r=r$_*$) = 4.6 10$^{11}$
M$_{\odot}$. Note that Olling and Merrifield (1998) advocate lower
values for the distance of the Sun to the Galactic Center and
for the rotation velocity at the distance of the Sun. Recent
estimates of the escape speed from HIPPARCOS data are discussed
in Meillon et al. (1997), and fall within a range of
400 to 550 km s$^{-1}$.

\subsection{Disk heating and massive black holes}

The phenomenon of disk heating, i.e. the increase in velocity
dispersion of stellar populations as function of age is well known
for the solar neighbourhood. Stars are thought to be borne with a
velocity dispersion similar to the gas, i.e. 10 km s$^{-1}$. Due
to scattering of the stars in the fluctuations of the galactic
potential, the peculiar velocities of the star with respect to the
circular velocity increase, and an ensemble of stars of a given age 
has thus a higher velocity dispersion with increasing age. 

Several mechanisms have been proposed for the disk heating : 1)
massive black holes as being the constituents of dark matter halos
2) giant molecular clouds, and 3) shearing
bits and pieces of spiral arms. Predictions for the shape of the
velocity ellipsoid (the normalized length of the velocity vectors
in the three principal directions) have been made by Lacey 
(1984, 1991) for each of these three scenario's. The new HIPPARCOS 
results will certainly lead to a more profound examination of
this problem for the solar neighbourhood (e.g. Gomez et al.
1997, Dehnen \& Binney 1998).

Lacey \& Ostriker (1985) analyse the effect of massive black holes, 
assuming they are the only constituent of the dark halo, on the 
dynamical heating of the disk. They derive an upper limit of about 
10$^6$ M$_{\odot}$ for our Galaxy. Rix \& Lake (1993) point 
out that for small Sc galaxies this upper
limit will have to be lowered to 10$^4$
M$_{\odot}$, since their potential well is much shallower. Very
heavy black holes could even wreck the fragile disk of those
galaxies altogether. In view of this effect, most people exclude
massive black holes as an important constituent of dark matter.

\subsection{Improved local disk model}

Cr\'ez\'e et al. (1998) analyzed the new HIPPARCOS data and the
associated Coravel radial velocity databases, and created a proper
sample of stars for which they could reanalyze the galactic force
law in the z-direction. Their best solution is $\rho_{0}$ = 0.076
$\pm$ 0.015 M$_{\odot}$ pc$^{-3}$, which does not leave any
room for disk dark matter.

This result has consequences for an idea of Pfenniger \& Combes
(1994), i.e. that the dark matter is in the form of cold gas,
which is almost undetectable due to its fractal structure.
This idea is partly designed to explain the evolutionary sequence
from Sd (gas rich, dark matter important) to Sa galaxies (gas
poor, and apparently less dark matter within the optical radius)
(cf. Pfenniger, Combes \& Martinet 1994). It also
explains the coincidence noted by Bosma (1978) that the ratio of
total mass to gas mass surface density becomes constant in the
outer parts. Carignan et al. (1990) restated that result by noting
the similarity in shape between the computed rotation curve for the 
HI component with that of the dark halo component.

For the solar neighbourhood, the rotation curve for the disk 
component has to rise to a least about 180 km s$^{-1}$ to make a 
roughly flat total rotation curve with bulge and disk alone. 
For an exponential disk which
places the Sun roughly at the position of turnover 
of the rotation curve of the disk component ($\sim$ 2.2
times the scalelength), this corresponds to a surface density of
about 100 M$_{\odot}$ pc$^{-2}$. Gould, Bahcall \& Flynn (1996)
evaluate the ``visible'' components as follows : 13 M$_{\odot}$
pc$^{-2}$ due to the (ordinary) gas, 14 M$_{\odot}$ pc$^{-2}$
due to young stars, and 12 M$_{\odot}$ pc$^{-2}$ due to dwarfs.
Thus if all matter is in a thin disk, about 60 M$_{\odot}$ pc$^{-2}$
is ``missing''. If this matter is distributed as cold gas in a disk, 
it is hard to see why such a gas rich disk does not go unstable and
forms stars. Gerhard \& Silk (1996) propose instead that such matter
is distributed in a flattened halo.

\subsection{Microlensing results}

The recent results from the micro-lensing surveys have led to 
the construction of new Galactic mass models, with emphasis on
their capability to predict the microlensing rate. The results
from several years of the MACHO survey (Alcock et al. 1997), combined
with the EROS results (Renault et al. 1998), leave little room for 
compact objects in the galactic halo with masses in the range of 
10$^{-7}$ to 10$^{-3}$ M$_{\odot}$ (cf. Alcock et al. 1998).
The most likely mass of the handful events detected in
the direction of the Large Magellanic Cloud is about 0.5 M$_{\odot}$,
which leads to a debate about the real location of the
objects giving rise to the lensing phenomenon : they could belong 
to an outlying tidal streamer of the LMC (cf. Zhao 1998). 
Analysis of mass models taking into account a number
of dynamical contraints leads to the result that only a fraction
of the mass in the dark halo of the Galaxy could be made up by 
MACHO's (cf. Alcock et al. 1998).

\subsection{Satellites of the Milky Way}

These galaxies are excellent tools for studying the dark matter
problem. Not only do their movement around the Galaxy lead to
estimates of the total mass of the Galaxy or the Local Group,
their internal dynamics show that they themselves are probably
dominated by the dark matter.

Lin \& Lynden-Bell (1982) studied the dynamics of the Magellanic
Stream, and find that an extended dark halo is needed to model
this complicated system. Lynden-Bell (1994) emphasizes the coincidence
between the location of the Draco and Ursa Minor dwarf spheroidals and
the Magellanic Stream along a great circle, as well as the possible
existence of an older Fornax-Leo-Sculptor stream which could betray
the orbital path of larger satellites. The recent HI map of Putnam
et al. (1998) of the Magellanic system, with the detection of a 
possible leading stream, calls for a new study of the interaction
between the Magellanic Clouds and our Galaxy, in order to get more
insight in the extended dark halo around our Galaxy. 

\subsection{Dwarf galaxy halos}

The phase space evolution of dark matter can put a constraint on the
nature of it, as follows (cf. Tremaine \& Gunn 1979) : it can
be shown that there is a minimum mass for neutrino's which they
should have if they are to constitute the dark matter in dwarf
galaxies. This has spurred the quest for a thorough study of
dwarf galaxy dynamics. Not only dwarf spheroidals can give an
answer, also gas rich dwarfs can help here. The ability of
dark matter candidates to cluster or not on small scales has
led to the important distinction of cold dark matter (CDM), which
can cluster on small scales, and hot dark matter (HDM), which cannot,
but might be important e.g. on the scales of clusters and
superclusters. (cf. Blumenthal et al. 1984).

For dwarf spheroidals, kinematics of individual bright stars allows
estimates of the mass-to-light ratios. A recent update (Mateo 1994)
shows that for the Draco and Ursa Minor dwarf spheroidals, the 
mass-to-light ratio is about 100. Better surface photometry is now
available from the work of Irwin \& Hadzidimitriou (1995), and more
velocities are being collected for individual stars in all dwarf
spheroidals orbiting our Galaxy, with ever greater accuracy
(cf. Olszewski 1998). The conclusion that some of these systems have
a high central density of dark matter (as high as 1 M$_{\odot}$
pc$^{-3}$ for the Ursa Minor dwarf) seems fairly secure, although
tidal effects from the Galaxy remain an important source of
uncertainty.

For gas rich dwarfs, 21-cm HI line studies using the WSRT or the
VLA, have yielded results for e.g. DDO 154 (Carignan \& Freeman 1985),
DDO 170 (Lake, Schommer \& Van Gorkom 1990), NGC 3109 (Jobin \& 
Carignan 1990) and DDO 105 (Broeils 1992). All these galaxies are dark
matter dominated, and their dark halos, when modeled as isothermal 
spheres, or with Hernquist profiles, appear to be concentrated enough 
to exclude massive neutrinos as their constituents. Recent work by De 
Blok and his collaborators (e.g. De Blok \& McGaugh 1997) on low
surface brightness late type disk galaxies leads to a similar conclusion.

However, a new problem arises related to numerical simulations of 
cosmological models. Moore (1994) and also Navarro, Frenk \& White (1996)
find core radii for dwarf galaxies from their numerical simulations,
which are even smaller than those observed in the above mentioned 
four systems, although Kravtsov et al. (1998) claim to have solved
this problem. In any case, the rotation curve data for dwarf galaxies 
can be used to constrain cosmological models via such simulations, 
and it may well be possible that a standard CDM
$\Omega$ = 1 model is ruled out by them (e.g. Navarro 1998).

\section{Dark matter in spiral galaxies}

We saw already in section 1 how the study of extended galaxy rotation
curves showed that dark matter is needed to explain their shape in
the outer parts of spiral galaxies. Here we will concentrate on
modern work, which still has not succeeded in determining whether
the inner parts of galaxies are dominated by dark matter or not.

\subsection{Maximum disks ?}

It is customary to construct composite disk/halo mass models of
spirals assuming a ``maximum disk'' solution, or to adopt a ``best
fit''. In such models, the data from surface photometry are used,
assuming a constant mass-to-light ratio, to calculate the expected
rotation curve for the visible components, bulge and disk. From the 
observed HI gas density, and a suitable factor to include helium, 
a rotation curve is calculated also, and quadratically added to the
first one. The resulting curve is then compared with the observed
rotation curve, and an additional dark halo component is introduced
when necessary. For extended HI rotation curves, such an analysis
has been done by several authors, e.g. Begeman, Broeils \& Sanders
(1991). The constancy of the mass-to-light with radius is usually
justified by the absence of colour gradients, indeed, data
of De Jong (1996) shows that colour gradients are small, and if
present, in the sense that disks become bluer outwards. In the latter 
case the use of near infrared data is preferred, since it accounts 
better for the contribution to the mass of the old stellar disk.

Athanassoula et al. (1987) introduce criteria from spiral
structure theory, and in particular those for swing amplification 
(Toomre 1981),
in order to get limits on the dynamical importance of the disk.
This leads to a range of values possible for (M/L)$_{\sf disk}$, in 
case one is asking for the possibility to have m = 2 structures
and for the suppression of m = 1 structures. They find that the
requirement to have halos with non hollow cores usually is consistent
with the absence of m = 1 structures, and that such models are
preferred when considerations of stellar populations and the buildup
of Sc disks at a constant rate of star formation over a Hubble time
are taken into account.

Bottema (1993), from an analysis of velocity dispersions, claims
that the maximum velocity of the disk component is 63\% of the
maximum observed velocity. The path to this result is strewn with
assumptions, the most important of which are that disks are
exponential with a velocity ellipsoid close to that in the solar
neighbourhood, that Freeman's (1970) law holds, and that
(B-V)$_{\sf old~disk}$ = 0.7 for all disks.
For NGC 3198 his result corresponds closely to the ``no m = 2''
solution proposed by Athanassoula et al. (1987). 

Recent work by Navarro (1998), based on fitting rotation curves
with a dark halo profile which fits well the cosmological simulations
of Navarro, Frenk \& White (1996), show that in his mass models
the dark matter also dominates in the inner parts of spiral galaxies.
Indeed, his decompositions for NGC 3198 are so dark matter dominated
that m = 2 structures will not be swing amplified at any radius.

However, Debattista \& Sellwood (1998) produce a clear argument
in favour of maximum disks : the dynamical friction of a bar against
a dark halo slows it down, and only in a maximum disk situation does
the corotation radius at the end of the simulation
extend to roughly 20\% further than the end
of the bar. If the halo is more concentrated, as in Navarro's models,
the bar slow down is so strong that corotation is at several times
the bar length, completely inconsistent with current notions about
bar pattern speeds. From realistic hydrodynamical simulations of the
gas flow in barred spirals, which well mimic the observed dust lanes
as regions of strong shocks, Athanassoula (1992) places corotation at
about 1.2 $\pm$ 0.2 times the bar length. Other determinations of
bar pattern speeds, based e.g. on the location of rings, which are
presumed to be linked to resonances, concur with this (e.g. Elmegreen
1996 for a review).

\subsection{Declining rotation curves}

Some galaxies have rotation curves which decline just beyond the
optical image, and stay more or less flat thereafter. Early examples
are NGC 5033 and NGC 5055 in Bosma (1978, 1981a), and also NGC 5908
(Van Moorsel 1982). Two more examples, NGC 2683 and NGC 3521, were 
given by Casertano \& Van Gorkom (1991), who speculated that declining
curves are linked with disks having short scalelengths. However,
Broeils (1992) finds cases of declining curves for galaxies with large
disk scalelengths. 

Declining rotation curves, because of the additional identifiable 
feature in the rotation curve, might hold out a promise to enable
us to discriminate between the various mass models. Since one
expects them to be found amongst galaxies with high
rotational velocities, I made a small survey with the VLA 
of a number of galaxies with W$_{\rm R}$ $>$ 400 km/s 
in collaboration with Van Gorkom, Gunn,
Knapp and Athanassoula. Several new cases of galaxies with declining
rotation curves were found. In Bosma (1998) a preliminary account is 
given for the most spectacular case, NGC 4414, for which also radial 
velocities and velocity dispersion information was obtained.

Unfortunately, the range in disk mass-to-light ratios for that galaxy
cannot be constrained very easily, in spite of the feature. However,
the velocity dispersion data allow the evaluation of the Toomre 
Q-parameter, which is found to be about 1.1 for a maximum disk model,
but 2.3 for a ``no m = 2'' model. The latter value is definitely too
high to allow spiral structure from swing amplification. 
A weak global spiral pattern is present in the old disk (Thornley
1996). Therefore, it seems unlikely that the inner parts of
bright disk galaxies are dark matter dominated.

\subsection{Warps}

As already noted by Kahn \& Woltjer (1959), any non-spherical mass 
distribution will cause a differential precession of a warp shape, and
thus leads to an increasing corrugation of the outer disk. The
observed statistics
of warps are such that at least 50\% of all spirals are thought to 
be warped (cf. Bosma 1991). To explain the frequency of warped HI
disks, it is thus necessary to have recourse to a mechanism which can 
keep them going. Several proposals have been made, none of them
entirely satisfactory (cf. Binney 1992).

Bosma (1991) also shows that the frequency of warps depends on the
ratio of halo core radius to optical radius of the galaxy : galaxies
for which this ratio is small do not have warped HI disks. This is
usually attributed to dynamical friction between a misaligned disk
and a dark halo. If the dark halo is strongly concentrated, such
misalignments are short lived, as is shown also by numerical
simulations (Dubinski \& Kuijken 1995).

\subsection{Shape of dark matter halos}

\subsubsection{Polar ring galaxies}

A special class of galaxies are the polar ring galaxies. These are
small S0 galaxies seen edge-on, and have an additional ring (annulus)
of matter orbiting over the poles. A first study of one of them,
A0136 - 0801 (Schweizer, Whitmore \& Rubin
1983), indicated that the material over
the pole had roughly the same rotation speed as the stellar disk.
From this, it was concluded that the dark halo must be nearly
spherical. However, later studies using more accurate data, and
including proper modelling of the self gravity of the polar ring,
changed this conclusion. Sackett et al. (1994) show that for NGC 4650A
the dark matter halo is quite flattened, but new data for this galaxy
by Arnaboldi et al. (1997) and modelling by Combes \& Arnaboldi (1996)
shows that the situation is even more complicated yet.

\subsubsection{Axisymmetry of the disks}

Since the natural shape of dark matter is triaxial (Binney 1978),
as is confirmed as well from the cosmologicial N-body simulations,
it is surprising at first sight that disk galaxies are roughly
axisymmetric in the outer parts. 
Yet this is borne out even for very high quality HI
data. Schoenmakers (1998) analyzed data for several well 
studied spiral galaxies, and find a very high degree of axisymmetry.
The same conclusion can be drawn from the work of Rix \& Zaritsky
(1995) on face-on spirals. This means that the process of disk
galaxy formation is such that the original triaxial dark matter
halo shape is modified by the formation of the disk, e.g. due to
dissipation (cf. Dubinski 1994). In any case the shape of the dark
matter halo close to the disk should be either oblate or prolate.

\subsubsection{Flaring gas layers}

A direct way to study the vertical shape of the dark halo is using
the variation of the thickness of the gas layer with radius in
edge-on galaxies. Predictions for this can be modelled quite 
straightforwardly from multicomponent disk/halo mass models (e.g. 
Athanassoula \& Bosma 1988, Bosma 1994), but in practice the effects 
of angular resolution and sensitivity, small warps, lopsidedness,
residual inclination effects, etc. may make
the observational determination difficult. Moreover, the problem is
directly dependent on the assumed value of the velocity dispersion
of the gas. Even so, Olling (1996ab) determined for the edge-on galaxy
NGC 4244 that its dark matter halo must be quite flattened, like an
E5 - E9 shape. This result contrasts with the results of Hofner \&
Sparke (1994), who found much rounder halos, based on their
interpretation of the warping behavior of the HI disks of several
inclined galaxies.

\subsection{Tidal tail extent}

In their survey of galactic bridges and tails,
Toomre \& Toomre (1972) succeeded in
producing numerical models of tidal encounters between galaxies which,
for a given moment in the evolution sequence, and using the information
on the spatial orientation, resembles closely the observed system
such a the M51 system, the Antennae (NGC 4038/39), etc. The basic
postulate in such models is that gravity is the dominant force (it
was frequently thought before that that bridges and tails needed to 
be explained by invoking magnetic forces).

As a corollary to this work, it was found that for slow encounters 
between spirals the objects merge to form an elliptical like object,
and Toomre (1977) produced a sequence of peculiar galaxies
which he considered in various stages of a merging process. This
merger hypothesis proves very pervasive, and now forms an integral
part of the ``bottom-up'' scenario's of galaxy and structure
formation. A full observational study
of the sequence of galaxies discussed by Toomre (1977) can be found
in Hibbard \& Van Gorkom (1996).

Since the study of Toomre \& Toomre (1972) 
was done without dark halos, it was
interesting to see whether the addition of halos to N-body simulations
would change the story. Barnes (1988) did this
for the Antennae (NGC 4038/39), with success. The growing capablities of
supercomputers to address the gravitational N-body problem led
Dubinski, Mihos \& Hernquist (1996) to reexamine the problem, with the
specific intent to delimit the extent of the dark halo. They thus 
considered models with different halo extent, and found that for 
large halo to disk mass tail forming is inhibited. Dubinski, 
Hernquist \& Mihos (1998) extended
these calculations, in order to establish a clear criterion to 
delimit halo extent. However, Barnes (1998) shows that their
conclusions are not true for haloes with very shallow density
profiles, and thus tidal tail extent may not be a helpful indicator
which can be used to rule out very large halos (see also Springel
\& White 1998).

\subsection{Extended dark halos}

The evidence for very extended dark halos around spiral galaxies
does not come from a single tracer such as HI. Other tracers are
brought to bear, such as satellites, binary galaxy statistics,
etc. The use of these tracers is much less straightforward, and
needs the development of mass estimators, which take into account
the statistical effects of the orbits of the tracer galaxy around 
the parent galaxy.

An example of this is the work by Zaritsky \& White (1994) and
Zaritsky et al. (1997) on a sample of spiral galaxies with dwarf 
satellites. These authors have been
slowly collecting data on dwarfs around large, inclined spirals,
in order to put together a well defined sample in which the satellite
distribution around the primaries can be treated statistically. They
conclude that large massive halos, with radii as large as 200 kpc,
do indeed exist around large spirals. However, the statistics of
the satellite distribution show some interesting details, which
are not fully understood in the framework of current ideas on galaxy
formation and evolution.

A completely different way to probe the extent of dark halos around
individual galaxies comes from the analysis of weak but measurable
changes to the shapes of distant galaxies due to the gravitational
lensing by foreground galaxies. Brainerd et al. (1996) report on
one such analysis, and find that large halos, of $\sim$ 100 h$^{-1}$ kpc
do indeed exist, in agreement with the conclusions from Zaritsky
\& White (1994).

\section{Dark Matter in other systems, and at larger scales}

I will treat these subjects only briefly, to show mainly that the
previous chapters are only the tip of the iceberg as far as the
dark matter problem is concerned. For more information one can
consult general cosmology textbooks, and Binney \& Tremaine (1987).
This subject is developping very rapidly, so I have selected some
highlights only.

\subsection{Dark Matter in Ellipticals}

For ellipticals, it has been much more difficult to establish that
they too are imbedded in a dark matter halo. This is due to the fact
that their luminosity distribution drops off very steeply with radius,
and that they, ordinarily, do not contain neutral hydrogen. Moreover,
the interpretation of the stellar velocity
data in the outer parts depends also on
the anisotropy of the velocity dispersion tensor.
For a few systems which do have neutral hydrogen, the data show generally
a flat rotation curve, hence a dark halo is inferred. The best studied
system this way is IC 2006, which has an outer ring of HI. Indications
are that the halo around this system is close to axisymmetric
(cf. Franx, Van Gorkom \& De Zeeuw 1994).

Recent efforts in detailed modelling of high quality
stellar radial velocity and velocity dispersion data have resulted
in the demonstration that dark matter is a necessary ingredient to
get good fits of the models to the data. This was done for NGC 2434 
(Rix et al. 1997) and NGC 6703 (Gerhard et al. 1998).
 
Other tracers can be used. A fruitful one comes from the
kinematics of planetary nebulae. The detection of such objects around
individual galaxies have now become routine and for a few ellipticals
the data indicate the presence of dark mass at large radii. This
has been shown in particular for NGC 3384 (Tremblay et al. 1995) and
NGC 5128, also known as Centaurus A (Hui et al. 1995). For this last
system, the dynamics of the outer shells detected in neutral hydrogen
also indicate the presence of a dark halo (Schiminovich et al. 1994).

The detection of X-ray gas in and around elliptical galaxies in clusters
has lead early on to the detection of large dark matter halos around
ellipticals like M87 and NGC 1399, which are in the center of
clusters. For normal ellipticals, only recent data confirm the presence
of extended dark matter halos around them. In particular, Buote \&
Canizares (1998) use a geometrical test to show the presence of dark
matter around three field ellipticals.

\subsection{Groups and clusters}

The traditional way to demonstrate the presence of dark matter has
been to collect radial velocity data from the individual members of
a group and cluster, from optical and/or radio data, and to apply
some form of the virial theorem. This has been pioneered by Zwicky 
(1933), and became popular in the 70s and 80s. Data for many groups
and clusters indicate high mass-to-light ratios, of order 100 - 500,
which is much higher than expected for the mass-to-light ratios of
individual galaxies which are of order 5 - 10 if they do not contain
much dark matter.

Another way to study dark matter comes from X-ray data, assuming 
hydrostatic equilibrium. The enclosed mass within a given radius
depends on the temperature of the hot gas giving rise to the X-ray
emission, its radial gradient, and the gradient
of the gas density. Mapping the latter using X-ray imaging is
rather straightforward, but the determination of the temperature
gradient is rather more difficult. In the 80s, data of the
Einstein satellite were used to determine estimates of the
dark matter content of several clusters of galaxies. Further
improvement of the data came from the Rosat and ASCA
satellites, and soon high quality data will come from the
AXAF and XMM facilities.

A third way to determine masses of clusters is using gravitational 
lensing, using arcs and arclets. This has grown from the first
demonstration of that arcs are due to lensing (Soucail et al. 1988)
to an impressive field in its own right, in particular with imaging
data from the Hubble Space Telescope, and the development of reliable
estimates of the mass of the lensing object from the distorted 
shapes of more distant galaxies.

An interesting study comparing all three methods for a number of distant
clusters is the one by Smail et al. (1996). They find reasonable
agreement between the results from the X-ray and lensing data, but
the optical data do not correspond that well, presumably due to the
influence of both substructure and interlopers in the samples of
galaxies used to determine the cluster velocity dispersion. In any
case, they confirm the high mass-to-light ratios found previously,
and give an upper bound to $\Omega$ of $\sim$ 0.4.

Compact groups of galaxies are a somewhat special case of ordinary
groups, but their properties are quite interesting (see Hickson 1997
for a review).
If there is little dark matter in a common halo around such groups,
numerical simulations show that the galaxies should merge quite fast
into one object. However, inclusion of a large extended common halo,
such a the one found for HCG 62 from X-ray data, will retard the
merging timescale to longer than a Hubble time (cf. Athanassoula,
Makino \& Bosma 1997). X-ray observations of sparser groups also indicate
high mass-to-light ratios, and thus the presence of dark matter in
them.

\subsection{Dark matter in the Universe}

\subsubsection{Large scale structure}

Work on the distribution of galaxies has shown the presence of large
scale structure in the Universe. Large scale flows exist between these
structures, and indicate the presence of dark matter on ever larger
scales. A review of this field has been made by Dekel (1994). From
a variety of methods, one can infer that in general $\Omega$ stays
less than 1, but at least as large as 0.2 - 0.3.

\subsubsection{Big Bang nucleosynthesis and the amount of baryons}

One of the colloraries of the standard Big Bang model is the
calculation of the nucleosynthesis of the primordial elements.
A first calculation was done in the seventies, and compared
with the observational data on D$^2$, He$^3$, He$^4$ and Li$^7$.
These calculations and comparison data are now more and more refined,
but recent work still has it that the upper limit to baryonic
dark matter is $\Omega_{\sf B}$ $\leq$ 0.02 h$^{-2}$ with h $\equiv$
H$_{\sf 0}$/(100 km s$^{-1}$ Mpc$^{-1}$). (e.g. Copi, Schramm \& Turner
1995). The difficulties surrounding the determination of the Hubble
Constant need not to be emphasized, but a lower limit for h is 0.4,
and more likely h $\simeq$ 0.75 (Madore et al. 1998).
Thus $\Omega_{\sf B}$ is at most 0.13, and more likely about 0.04,
which is lower than the values observed in groups and clusters if
those were typical for the Universe as a whole. It is interesting to 
note that for some clusters, like the Coma cluster, the gas fraction 
as detected by X-rays alone might exceed the upper bound derived 
from Big Bang nucleosynthesis if the Universe is at closure density
(cf. White et al. 1993). 

\subsubsection{$\Omega$ = 1 ?}

In the early 80's, the most popular scenario for cosmology is the
inflation scenario, which could account for the fact that the observed
matter density in the universe is close to the critical density. In
fact, it was postulated that the matter density, expressed in terms
of the critical density, $\Omega$, is exactly 1. Thus, compared to the
results from Big Bang nucleosynthesis, there is a lot of dark matter
not in baryonic form, but most likely in the form of a Weakly
Interacting Massive Particle (WIMP). This hypothesis still persists
until today, since it is theoretically very attractive. However,
the actual measurements of the matter density, though difficult, come 
out to be $\Omega$ $\sim$ 0.2 (cf. Bahcall, Lubin \& Dorman 1995), which
lead some people to think that all of the dark matter could still
be baryonic. Only further work will tell what the real answer is.

\section{Concluding remarks}

From this brief review, one can see that the study of the local
Universe - our Galaxy, its near neighbours, and nearby galaxies
in general - provides many contraints on the dark matter problem,
including on what it actually may consist of. The study of the 
more distant Universe is complementary to this, and though the
next generation of large telescopes should provide a lot more 
information on it, only a combined approach will solve the entire
dark matter problem.

\section*{acknowledgements}

I would like to thank Lia Athanassoula for her comments on the
manuscript while it was being written, and for frequent discussions
of the dark matter problem in general.

\end{document}